\documentclass[a4paper,11pt]{article}

\usepackage{amsmath}
\usepackage{listings}
\usepackage{url}
\usepackage{graphicx}
\usepackage{fullpage}

\newcommand{\ie}{\emph{ie.\/}}

\newcommand{\ket}[1]{\ensuremath{|#1\rangle}}
\newcommand{\bra}[1]{\ensuremath{\langle#1|}}

\newcommand{\braket}[2]{\ensuremath{\langle{#1}|{#2}\rangle}}

\newtheorem{definition}{Definition}
\newtheorem{example}{Example}
\newcommand{\keywords}{\textbf{Keywords:} }
\title{Separable and non-separable data representation for pattern discrimination}

\author{Jaros{\l}aw Adam Miszczak\\
IITiS PAN\\ Baltycka 5, 44100 Gliwice, Poland\\ 
\quad 
\\ 
University of Cagliari,\\ Via~Is Mirrionis 1, 09123 Cagliari, Italy\\
\texttt{miszczak@iitis.pl}}

\date{15/03/2015 (v. 0.35)}

\begin{document}

\maketitle

\begin{abstract}
We provide a complete work-flow, based on the language of quantum information
theory, suitable for processing data for the purpose of pattern recognition. The
main advantage of the introduced scheme is that it can be easily implemented and
applied to process real-world data using modest computation resources. At the
same time it can be used to investigate the difference in the pattern
recognition resulting from the utilization of the tensor product structure of
the space of quantum states. We illustrate this difference by providing a simple
example based on the classification of 2D data. 

%We use standard pattern classification algorithms translated into the language
%of quantum information theory and study the impact of the entanglement in the
%states used for the purpose of data representation.

\end{abstract}

%\tableofcontents

%%%%%%%%%%%%%%%%%%%%%%%%%%%%%%%%%%%%%%%%%%%%%%%%%%%%%%%%%%%%%%%%%%%%%%%%%%%%%%%%
\section{Introduction}
%%%%%%%%%%%%%%%%%%%%%%%%%%%%%%%%%%%%%%%%%%%%%%%%%%%%%%%%%%%%%%%%%%%%%%%%%%%%%%%%
Quantum machine learning aims at merging the methods from quantum information
processing and computer engineering to provide new solutions for problems in the
areas of pattern recognition and image
understanding~\cite{schuld14introduction,wittek14quantum,wiebe15quantum}. From
one side the research in this area is focused at applying the methods of quantum
information processing for solving problems related to classification and
clustering in signal processing. From the other perspective the methods for
classification developed in computer engineering are used to find solutions for
problems like quantum state
discrimination~\cite{helstrom,chefles00quantum,hayashi05quantum,markham08quantum},
which are tightly connected with the recent developments in quantum
cryptography.

The aim of the paper is to present a complete work-flow, based on the
mathematical formalism used in quantum information theory, suitable for
processing data for the purpose of pattern recognition. The main goal is to
introduce the method for describing classical data using entangled quantum
states and to investigate the difference between the entangled and the separable
representations of states. The presented work-flow enable the processing of
real-world data as it can be implemented and executed on standard desktop
computers. The main reason why this is possible is the introduction of the
quantization procedure which allows the reduction of the dimensionality of the
utilized quantum states.

This paper is organized as follows.
In Section~\ref{sec:preliminaries} we provide some notation and initial
considerations related to the problem of using quantum states for the purpose of
pattern representation and classification. 
In Section~\ref{sec:quantum-representation} we propose a general framework for
encoding data using the formalism of quantum states and describe the basic
classification of the 1D data using this representation. 
In Section~\ref{sec:higher-dims} we introduce a formalism for constructing
representations of feature vectors and provide some examples of using entangled
and separable flavors of the introduced formalism in 2D case.
Finally, in Section~\ref{sec:final} we provide a summary of the presented work
and provide concluding remarks.

%%%%%%%%%%%%%%%%%%%%%%%%%%%%%%%%%%%%%%%%%%%%%%%%%%%%%%%%%%%%%%%%%%%%%%%%%%%%%%%%
\section{Preliminaries}\label{sec:preliminaries}
%%%%%%%%%%%%%%%%%%%%%%%%%%%%%%%%%%%%%%%%%%%%%%%%%%%%%%%%%%%%%%%%%%%%%%%%%%%%%%%%
We start by introducing the notation from the area of quantum information
relevant for the purpose of processing patterns. We also review the problem of
pattern classification and introduce the notation used in the paper.

%%%%%%%%%%%%%%%%%%%%%%%%%%%%%%%%%%%%%%%%%%%%%%%%%%%%%%%%%%%%%%%%%%%%%%%%%%%%%%%%
\subsection{Kets, tensors and features}
%%%%%%%%%%%%%%%%%%%%%%%%%%%%%%%%%%%%%%%%%%%%%%%%%%%%%%%%%%%%%%%%%%%%%%%%%%%%%%%%

In this paper we use the notation of quantum mechanics in which vectors
are written using ket symbols, such that
\begin{equation}
\ket{a_1}
\end{equation}
is a base vector with label $a_1$. In the simplest case of two-dimensional
space, which can be used to represent a bit on a quantum computer, we have
\begin{equation}
\ket{0} \equiv \binom{1}{0} \quad \mathrm{and} \quad \ket{1} \equiv \binom{0}{1}.
\end{equation}
Accordingly, for $N$-dimensional space we have 
\begin{equation}
\ket{i} \equiv \left(
\begin{smallmatrix}
0\\ \vdots \\ 1 \\ \vdots \\ 0
\end{smallmatrix}
\right)
\begin{smallmatrix}
  \\ \leftarrow i^\mathrm{th}\ \mathrm{position,} \\ \\
\end{smallmatrix}
\end{equation}
with $i\in\{0,1,\dots,N-1\}$. We also assume that ket vectors are normalized.

This notation can be used to express any combination of vectors as
\begin{equation}
a\equiv\ket{a} = \sum_{i}a_i\ket{i}.
\end{equation}
Usually we require that $\sum_i|a_i|^2 =1$.

Let us now assume that we deal with $p$ features, which are measured in the
experiment. With each feature we connect a vector space $F_i$ of dimension
$K_i$, necessary to store the possible measurement results.

A pattern consisting of $p$ features can be represented as a tensor product of
feature spaces as
\begin{equation}
F_1\otimes F_1\otimes \dots \otimes F_p,
\end{equation}
where $\otimes$ denotes tensor (a.k.a. Kronecker)
product~\cite{vanloam00ubiquitous}. For two vectors $\ket{a}\in F_1$ and
$\ket{b}\in F_2$ 
\begin{equation}
\ket{a} = \left(
\begin{smallmatrix}
a_1\\
a_2\\
\vdots\\
a_{K_1}
\end{smallmatrix}
\right)
\quad \mathrm{and}\quad
\ket{b} = \left(
\begin{smallmatrix}
b_1\\
b_2\\
\vdots\\
b_{K_2}
\end{smallmatrix}
\right)
\end{equation}
tensor product is defined as
\begin{equation}
\ket{a}\otimes\ket{b} =\left(
\begin{smallmatrix}
a_1 b_1\\
%a_1 b_2\\
\vdots \\
a_1 b_{K_2}\\
\vdots \\
a_{K_1} b_1\\
%a_{K_1} b_2\\
\vdots \\
a_{K_1} b_{K_2}\\
\end{smallmatrix}
\right),
\end{equation}
and it is customary to write
$
\ket{a}\otimes \ket{b} \equiv \ket{a,b} \equiv \ket{ab}.
$

One of the features of the above formalism is the seamless notation of the
scalar product as
\begin{equation}
a\cdot b \equiv \braket{a}{b} = \sum_{i,j} a_i b_j \bra{i}  \ket{j}.
\end{equation}
Elements $\bra{i} \ket{j} = \delta_{ij}$ can be written as $\braket{i}{j}$ and
they represent scalar products of the base vectors and $\bra{a}$ represents the
vector dual to $\ket{a}$.

%%%%%%%%%%%%%%%%%%%%%%%%%%%%%%%%%%%%%%%%%%%%%%%%%%%%%%%%%%%%%%%%%%%%%%%%%%%%%%%%
\subsection{Pattern recognition}
%%%%%%%%%%%%%%%%%%%%%%%%%%%%%%%%%%%%%%%%%%%%%%%%%%%%%%%%%%%%%%%%%%%%%%%%%%%%%%%%
Let us now assume that we are dealing with the problem of the supervised
learning. In this situation we have at our disposal a number of representatives
associated with labels denoting classes. The classes used to discriminate are
constructed using a number of measurement results, each one representing a
pattern of a known class~\cite{webb,webb-copsey},
\begin{equation}\label{eqn:classical-class}
C_i = \{c_k^{(i) } : k=1,\dots,n_i\},
\end{equation}
where number of representatives $n_i$ can be different for each class. The set
of class elements with the associated labels is refereed to as the learning set.

The classification of the unknown patter $x$ is performed using the score
function $g(x,i)$, calculated for the pattern and each class. In the simplest
scenario, the score function is calculated as a distance between the unknown
pattern and the class representatives. The class representatives
$\rho^{(i)}_\mu$ for each class $i=1,2,\dots,C$ are built by the appropriate
averaging over the class elements.

The unknown pattern is classified according to the greatest value of the score
$q^{(i)}(x)$, calculated as the distance between the pattern $x$ and the class
representatives $\rho^{(i)}_\mu$, $i=1,2,\dots,C$,
\begin{equation}
q^{(i)}(x) = D(\rho^{(i)},x).
\end{equation}

%%%%%%%%%%%%%%%%%%%%%%%%%%%%%%%%%%%%%%%%%%%%%%%%%%%%%%%%%%%%%%%%%%%%%%%%%%%%%%%%
\section{Pattern representation in quantum formalism}\label{sec:quantum-representation}
%%%%%%%%%%%%%%%%%%%%%%%%%%%%%%%%%%%%%%%%%%%%%%%%%%%%%%%%%%%%%%%%%%%%%%%%%%%%%%%%
In this section we develop a quantum representation for patterns and classes. We
focus on the patterns distributed with the normal distribution, but the provided
framework does not depend on this assumption. We start with the simplest case of
1D data and subsequently we consider the possible generalizations of the
introduced formalism for the purpose of discrimination patterns described by
vectors of features.

%%%%%%%%%%%%%%%%%%%%%%%%%%%%%%%%%%%%%%%%%%%%%%%%%%%%%%%%%%%%%%%%%%%%%%%%%%%%%%%%
\subsection{General framework}
%%%%%%%%%%%%%%%%%%%%%%%%%%%%%%%%%%%%%%%%%%%%%%%%%%%%%%%%%%%%%%%%%%%%%%%%%%%%%%%%
In order to define a quantum classification procedure one needs to specify the
mapping between the space of values and the space of quantum states. This is
exactly the counterpart of the quantization function used in machine learning
algorithms. However, in this case, this function is used not only to reduce the
dimensionality of the problem, but also to obtain a quantum representation of
patterns and classes. 

Here we aim at representing data in the form of vectors --- pure quantum states
--- or corresponding projection operators. In principle one can extend the
presented formalism to include mixed quantum states, but the particular choice
of pure states enables the straightforward utilization of quantum superposition
and quantum entanglement.

%%%%%%%%%%%%%%%%%%%%%%%%%%%%%%%%%%%%%%%%%%%%%%%%%%%%%%%%%%%%%%%%%%%%%%%%%%%%%%%%
\subsubsection{Distance-based methods}
%%%%%%%%%%%%%%%%%%%%%%%%%%%%%%%%%%%%%%%%%%%%%%%%%%%%%%%%%%%%%%%%%%%%%%%%%%%%%%%%
From the quantum information point of view, the most natural formulation of the
pattern classification problem is based on the distance minimization.

In distance-based methods we are interested in the distance between the class
representative and the pattern we wish to classify~\cite{Cunningham2008}. In
order to use quantum states for the purpose of pattern classification we use a
framework based on the following assumptions:
\begin{itemize}
	\item[(1)] Patterns are represented as normalized vectors.
	\item[(2)] Classes are represented by the normalized superposition of the
	representatives.
	\item[(3)] Similarity between the patterns is calculated as a scalar product.
\end{itemize}

The first assumption from the above list requires mapping the classical data on
the quantum pure states,
\begin{equation}
x \mapsto \ket{f(x)}.
\end{equation}
In the following the procedure executing this step will be called the quantum
quantization and its goal is twofold. Firstly, it is required to reduce the
feature space and generate the quantization symbols. Secondly, it calculates the
dimensions of the state space used and maps the quantization symbols onto this
space. The quantum quantization step is crucial from the point of view of
analyzing the quantum correlations as it effectively defines the structure of
the used state space.

For a given quantum quantization procedure and a selected distance measure one
obtains a complete procedure for quantum-based classification. In order to
classify an unknown pattern $x$ one has to execute the following steps.
\newcounter{step}
\begin{list}{{\rm \bf Step \arabic{step}:}}{\usecounter{step}}
    \item Perform quantum quantization by mapping class elements 
	$$
	c^{(i)}_j \mapsto \ket{f(c^{(i)}_j)}.
	$$
    \item Prepare the class representatives $\rho^{(i)}_\mu$ for each class
    $i=1,2,\dots,C$ by the appropriate averaging over the quantum class
    elements.
	\item Calculate the distance between the pattern $x$ and the class
	representatives $\rho^{(i)}_\mu$, $i=1,2,\dots,C$,
	$$
	q^{(i)}(x) = D(\rho^{(i)},x).
	$$
	\item Classify the pattern according to the greatest value of $q^{(i)}(x)$.
\end{list}

One should note that the class representatives $\rho^{(i)}_\mu$, $i=1,2,\dots,C$
can be obtained as a flat superposition of the learning set elements. We will
use this method in the next part of the paper.

%%%%%%%%%%%%%%%%%%%%%%%%%%%%%%%%%%%%%%%%%%%%%%%%%%%%%%%%%%%%%%%%%%%%%%%%%%%%%%%%
\subsubsection{Nearest neighbors method}
%%%%%%%%%%%%%%%%%%%%%%%%%%%%%%%%%%%%%%%%%%%%%%%%%%%%%%%%%%%%%%%%%%%%%%%%%%%%%%%%
The quantum states obtained using the quantum quantization procedure can be used
in any scheme for pattern recognition. For example, in the case of k-NN scheme,
one needs to perform the following steps.

\begin{list}{{\rm \bf Step \arabic{step}:}}{\usecounter{step}}
    \item Perform quantum quantization by mapping class elements 
	$$
	c^{(i)}_j \mapsto \ket{f(c^{(i)}_j)}.
	$$
    \item Calculate the distances $q_j^{(i)}(x) = D(\ket{f(c^{(i)}_j)},x)$
    between the class representatives $\rho^{(i)}_\mu$ and the pattern for each
    class $i=1,2,\dots,C$.

	\item Classify the pattern according to the voting using values of
	$q_j^{(i)}(x)$ uilizing $k$ values.
\end{list}

However, in this case the procedure is reduced to the classical one as it does
not utilize the superposition of the base vectors. For this reason we restrict
our attention to the distance-base methods.

%%%%%%%%%%%%%%%%%%%%%%%%%%%%%%%%%%%%%%%%%%%%%%%%%%%%%%%%%%%%%%%%%%%%%%%%%%%%%%%%
\subsection{1D case with flat subspaces}\label{sec:representation-flat}
%%%%%%%%%%%%%%%%%%%%%%%%%%%%%%%%%%%%%%%%%%%%%%%%%%%%%%%%%%%%%%%%%%%%%%%%%%%%%%%%
In the following we focus on the features represented as numerical values.

We start by considering the 1D feature vectors and introducing the quantum
quantization procedure.  In this case features (and class elements) are
represented using flat representation in the subspace of the appropriate quantum
state space. In this context flat means that patterns are represented by first
transforming them into equally distributed intervals and next by representing
the intervals as quantum state vectors.

In orders to fix the dimension of the quantum state space required to encode the
quantization symbols, we will use the data from the learning set. We also
require that for the fixed state space of dimension $d$, vectors $\ket{0}$ and
$\ket{d-1}$ will encode all elements smaller and all elements larger than the
elements in the learning set. For this reason we required that in the case of
two classes
\begin{equation}
d = |[\max(C_1 \cup C_2)] - [\min(C_1 \cup C_2)]| + 2.
\end{equation}
Here $[ c ]$ is the nearest integer function, sometimes denoted as
$\mathrm{nint(c)}$, defined as
\begin{equation}
[ c ] = \left\{
\begin{array}{ll}
-1^{\mathrm{sgn}(c)} \lceil |c| \rceil, & \mathrm{if}\ \{c\}<1/2\\
-1^{\mathrm{sgn}(c)} \lfloor |c| \rfloor, & \mathrm{if}\ \{c\}\geq 1/2
\end{array}\right.,
\end{equation}
where $\{c\}$ denotes the integer part of $c$, $\{c\} = |\lfloor |c| \rfloor -
|c||$ This function is implemented as \texttt{rint} function in \texttt{math.h}
C standard library and as \texttt{Round} function in \emph{Mathematica} system. 

For a given number $c$ we construct its quantum representation as vector
\begin{equation} \label{eqn:num-to-vec}
q_{\mathrm{fs}}(c) = \ket{[ c ] - \{\min(C_1 \cup C_2)\} + 1 },
\end{equation}
in $d$-dimensional space of pure quantum states. 

The above procedure can be implemented in \emph{Mathematica} system as
\begin{lstlisting}
qMap[x_, ns1_, ns2_] := Block[{qMin, qMax, qDim},
  qMin = Round[Min[Join[ns1, ns2]]];
  qMax = Round[Max[Join[ns1, ns2]]]+1;
  qDim = Abs[qMax - qMin] + 2;
  Which[
   x < qMin, Ket[0, qDim],
   x > qMax, Ket[qDim - 1, qDim],
   True, Ket[Round[x] - qMin + 1, qDim]
   ]
  ]
\end{lstlisting}

Using the above quantization we define for each class its representation as
\begin{equation}
\ket{\mu_i}=\frac{\ket{\widetilde{\mu_i}}}{\| \ket{\widetilde{\mu_i}} \|},
\end{equation}
where $\ket{\widetilde{\mu_i}}=\sum_{k}q_{\mathrm{fs}}(c^{(i)}_k)$ represents an
unnormalized superposition of the elements representing the class.

Finally we introduce quantum flat subspace (QFS) classifier using the score
function
\begin{equation}
g_i^{QFS}(c) = |\braket{c}{\mu_i}|,
\end{equation}
for $i=1,2,\dots,n$ representing classes $C_i, C_2,\dots,C_n$.
Using this we classify pattern $c$ into class $i$ iff
\begin{equation}\label{eqn:qfs-classifier}
g_i^{QFS}(c) \geq g_j^{QFS}(c)
\end{equation}
for all $i\not=j$.

\begin{example}
Let us consider two samples of real numbers $C_1$ and $C_2$ distributed with
$\mathcal{N}(-2,1)$ and $\mathcal{N}(2,1)$ respectively.

For the purpose of classification we represent each class as a normalized
superposition of its representatives. For example if $C_1$ is represented by
\begin{equation}
\{-2.24697, -1.17115, -0.882941, -1.9828\}
\end{equation}
and $C_2$ by
\begin{equation}
\{0.836746, 1.70144, 3.05605, -0.0344292\}
\end{equation}
we get quantum representations for both classes as
\begin{equation}
\ket{\mu_1}=\frac{1}{\sqrt{6}}\left(\ket{0} + \ket{1} + 2 \ket{2}\right)
\end{equation}
and
\begin{equation}
\ket{\mu_2}=\frac{1}{2}\left(\ket{3}+ \ket{4} + \ket{5} + \ket{6}\right)
\end{equation}
respectively for $C_1$ and $C_2$. Both $\ket{\mu_1}$ and $\ket{\mu_2}$ are
normalized and each component has an amplitude representing a relative frequency
of the appropriate element in the learning set.

One should note that we need to choose the dimension of the subspace encoding
the integer part to be large enough for encoding all representatives. For this
reason the representation will require higher dimensional state space if the
learning set contains more elements. In this example the observations are
encoded in the space of dimension $d=7$.

This classification gives the following output
\begin{equation}
\{(-4, 1), (-3, 1), (-2, 1), (-1, 1), (0, 2), (1, 2), (2, 2), (3, 2), (4, 1)\}
\end{equation}
where each pair represents a number and the assigned class.

Note that number 4 has been classified as belonging to class 1 due to the lack
of the representative corresponding to this number in both classes. For this
reason $g_1^{QFS}(4)=g_2^{QFS}(4)=0$ and the classifier cannot decide about the
result solely on the data included in the quantum state. To avoid the
inconclusive results we assign the pattern to the arbitrarily chosen class.
\end{example}

%%%%%%%%%%%%%%%%%%%%%%%%%%%%%%%%%%%%%%%%%%%%%%%%%%%%%%%%%%%%%%%%%%%%%%%%%%%%%%%%
\section{Higher dimensional generalizations}\label{sec:higher-dims}
%%%%%%%%%%%%%%%%%%%%%%%%%%%%%%%%%%%%%%%%%%%%%%%%%%%%%%%%%%%%%%%%%%%%%%%%%%%%%%%%

In this section we provide methods for using the quantum encoded patterns
consisting of vectors of features. We introduce two basic schemes for encoding
patterns and classes -- separable and non-separable. Next, we provide some
examples of using these schemes. We also propose a hybrid method which only
partially exploits the tensor product structure of multipartite quantum states.

%%%%%%%%%%%%%%%%%%%%%%%%%%%%%%%%%%%%%%%%%%%%%%%%%%%%%%%%%%%%%%%%%%%%%%%%%%%%%%%%
\subsection{Quantum representations of feature vectors}
%%%%%%%%%%%%%%%%%%%%%%%%%%%%%%%%%%%%%%%%%%%%%%%%%%%%%%%%%%%%%%%%%%%%%%%%%%%%%%%%

\begin{figure}[ht]
	\centering
	\includegraphics[width=0.3\textwidth]{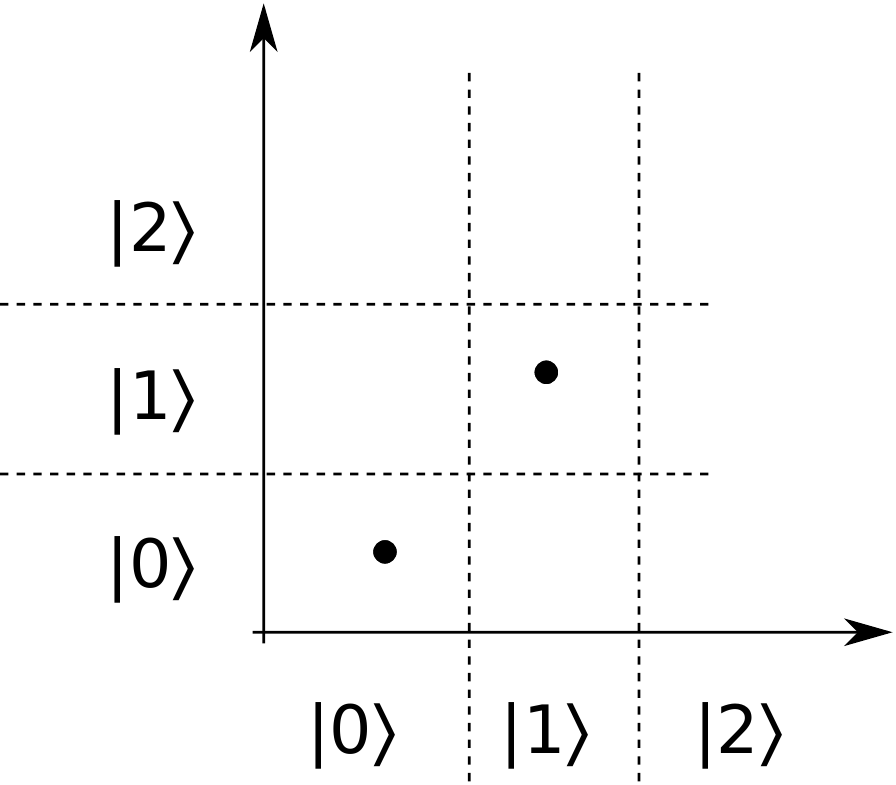}
	\caption{Learning set with two elements from the same class and the possible
	quantum quantization assigned to each coordinate.}
	\label{fig:ent_vs_sep}
\end{figure}

Let us use the learning set illustrated in Fig~\ref{fig:ent_vs_sep} as an
initial example. In this case the elements of the class are represented as
quantum states $\ket{00}$ and $\ket{11}$. We can use this representation and
construct the representation of the class as the vector
$\frac{1}{\sqrt{2}}(\ket{00}+\ket{11})$. In this situation we obtain the state
which is entangled. However, we can alternatively construct the representation
of the class in Fig.~\ref{fig:ent_vs_sep} component-wise and in this case we
obtain a pair of vectors resulting from the quantum averaging of the components,
namely $(\frac{1}{\sqrt{2}}(\ket{0}+\ket{1}),
\frac{1}{\sqrt{2}}(\ket{0}+\ket{1}))$. This can be translated into the
representation in the tensor product space as
$\frac{1}{2}(\ket{0}+\ket{1}+\ket{2}+\ket{3})$. 

%%%%%%%%%%%%%%%%%%%%%%%%%%%%%%%%%%%%%%%%%%%%%%%%%%%%%%%%%%%%%%%%%%%%%%%%%%%%%%%%
\subsection{Separable pattern representation}
%%%%%%%%%%%%%%%%%%%%%%%%%%%%%%%%%%%%%%%%%%%%%%%%%%%%%%%%%%%%%%%%%%%%%%%%%%%%%%%%
The most straightforward extension of the 1D example to the case of patterns
represented by vectors is obtained by using the 1D classifier independently from
each component. In this model the features are encoded into corresponding
quantum states and the classification is performed component-wise. As a result
one obtains a vector of results, which indicates the classification of the
pattern.

Following the example presented above, we introduce the separable data
representation as follows.

\begin{definition}[Separable data representation]
For a vector of features $x=(x_1,x_2,\dots,x_n)$ the quantum counterpart
obtained using the separable data representation is constructed as $n$-tuple of
quantum representations of the components, \ie
\begin{equation}
\ket{x} = \left(
\begin{smallmatrix}
\ket{x_1} \\
\ket{x_2}\\
\vdots \\
\ket{x_n}
\end{smallmatrix}
\right).
\end{equation}
\end{definition}

Accordingly, one can introduce the quantum averaging used to represent classes of
features as an averaging over the components. This can be expressed as follows. 

\begin{definition}[Separable class representation]
For a class consisting of $n$-dimensional features vector,
$C_i=\{c^{(i)}_1,c^{(i)}_2,\dots,c^{(i)}_k\}$, the quantum separable
representation of the class is obtained as
\begin{equation}\label{eqn:rep-class-separable}
\ket{\mu_i^S} = 
\frac{1}{\sqrt{k}}\left(
\begin{smallmatrix}
\sum_{j=1}^k \ket{c^{(i)}_{j,1}}\\
\sum_{j=1}^k \ket{c^{(i)}_{j,2}}\\
\vdots\\
\sum_{j=1}^k \ket{c^{(i)}_{j,n}}
\end{smallmatrix}
\right),
\end{equation}
where $c^{(i)}_{j,l}$ denotes the $l$-th element of the $j$-th feature vector in
the class $C_i$.
\end{definition}

In the simplest case of 2D feature space we have
\begin{equation}
\ket{(v_1,v_2)} = 
\left(
\begin{smallmatrix}
\ket{v_1}\\
\ket{v_2}
\end{smallmatrix}
\right)
\end{equation}
and the class representations are constructed as
\begin{equation}
\ket{\mu_k} = \frac{1}{\sqrt{2}}\left(
\begin{smallmatrix}
\sum_{i=1}^2\ket{v_{1,i}^k}\\
\sum_{i=1}^2\ket{v_{2,i}^k}
\end{smallmatrix}
\right).
\end{equation}

One should note that in the introduced scheme the Dirac notation is overloaded
and has to be interpreted according to the type of its argument. Moreover, the
quantum quantization procedure required to build the representations of
components has to operate on the quantum state space of dimension appropriate
for each feature. 

%%%%%%%%%%%%%%%%%%%%%%%%%%%%%%%%%%%%%%%%%%%%%%%%%%%%%%%%%%%%%%%%%%%%%%%%%%%%%%%%
\subsection{Non-separable pattern representation}
%%%%%%%%%%%%%%%%%%%%%%%%%%%%%%%%%%%%%%%%%%%%%%%%%%%%%%%%%%%%%%%%%%%%%%%%%%%%%%%%
One of the main differences between the classical and quantum mechanics is the
structure of the space used to describe the composite system. In the classical
case such space is constructed using the Cartesian product of the subspaces
describing the subsystem, whilst in the quantum case the composite systems are
described using the Kronecker product.

In quantum information theory the main phenomenon arising from the use of the
tensor product is the entanglement. As the entanglement can be seen as a form of
strongly non-classical correlation, we expect that the usage of the tensor
product structure allows us encoding and exploiting the correlation between the
features. For this reason we introduce the non-separable scheme as follows.

\begin{definition}[Non-separable data representation]
For a vector of features $x=(x_1,x_2,\dots,x_n)$ the quantum counterpart
obtained using the non-separable data representation is constructed as a tensor
product of quantum representations of the components, \ie
\begin{equation}
\ket{x} = 
\ket{x_1} \otimes \ket{x_2} \otimes \dots \otimes \ket{x_n}.
\end{equation}
\end{definition}

In the simplest case of 2D feature space we have
\begin{equation}
\ket{(v_1,v_2)} = \ket{v_1}\otimes\ket{v_2}.
\end{equation}

\begin{definition}[Non-separable data representation]
For a class consisting of $n$-dimensional features vector,
$C_i=\{c^{(i)}_1,c^{(i)}_2,\dots,c^{(i)}_k\}$, the quantum non-separable
representation of the class is obtained as

\begin{equation}
\ket{\mu_i^E}=\frac{1}{\sqrt{k}}\sum_k\ket{c_{k,1}^{(i)}}\otimes\ket{c_{k,2}^{(i)}}\otimes \dots \otimes\ket{c_{k,n}^{(i)}}.
\end{equation}
\end{definition}

The main difference between the above formula and the formula in
Eq.~(\ref{eqn:rep-class-separable}) is that it maps feature vectors onto vector
space of dimension $d_1\times d_2 \times \dots \times d_n$, where $d_i$,
$i=1,2,\dots,n$, are dimensions of spaces required to encode features. The
separable data representation requires only the space of dimension $\sum_i d_i$.

In the case of 2D feature space the above encoding can be implemented in
\emph{Mathematica} as
\begin{lstlisting}
qMuEnt2d[k_,ns_] := Block[{m1, m2, v},
  m1 = qMap[#, ns[1, 1], ns[2, 1]] & /@ ns[k, 1];
  m2 = qMap[#, ns[1, 2], ns[2, 2]] & /@ ns[k, 2];
  v = Plus @@ Flatten[Table[x $\otimes$ y, {x, m1}, {y, m2}], 1];
  v/Norm[v]
  ]
\end{lstlisting}
where function \lstinline/qMap/ was defined in Section~\ref{sec:representation-flat}.

%%%%%%%%%%%%%%%%%%%%%%%%%%%%%%%%%%%%%%%%%%%%%%%%%%%%%%%%%%%%%%%%%%%%%%%%%%%%%%%%
\subsection{Example: pattern recognition in 2D}
%%%%%%%%%%%%%%%%%%%%%%%%%%%%%%%%%%%%%%%%%%%%%%%%%%%%%%%%%%%%%%%%%%%%%%%%%%%%%%%%
The main advantage of the scheme introduced in this section is that it can be
easily implemented and executed for the real-world data. In this section we
present a simple example illustrating the difference between the separable and
the non-separable representation of data.

In order to execute the classification using the representations introduced
above, one has to specify the score function used in the final step. In the case
of the non-separable data representation we use the natural choice, namely the
overlap between the class representation and the quantum state representing the
pattern,
\begin{equation}
g^E(x,i) = |\braket{\mu_i^E}{x}|.
\end{equation}

In the case of the separable data representation we will use the score function
defined as
\begin{equation}
g^E(x,i) = \sqrt{\sum_k(\braket{\mu^S_{i,k}}{x_k})^2},
\end{equation}
where $\ket{\mu^S_{i,k}}$ denotes the $k$-th component of the separable quantum
representation defined in Eq.~(\ref{eqn:rep-class-separable}).

The above definitions allow the comparison of the efficiency of the pattern
classification. For this purpose we have used two 2D Gaussian sources producing
the pairs described as $(\mathcal{N}(1,1), \mathcal{N}(1,1))$ and
$(\mathcal{N}(-2,1), \mathcal{N}(-2,1))$. 

For the purpose of the presented experiment we have generated a sample of 100
learning sets, each consisting of $n$ elements of both classes, with
$n=2,3,\dots,16$. For each learning set we have used two samples of 1000 points
generated using the distribution describing the classes. Each of the generated
points has been classified using both separable and non-separable quantum data
representation. The test has been assessed as positive if the point had been
classified to the class corresponding to the appropriate distribution.

\begin{figure}[ht]
	\centering
	\includegraphics[width=0.55\textwidth]{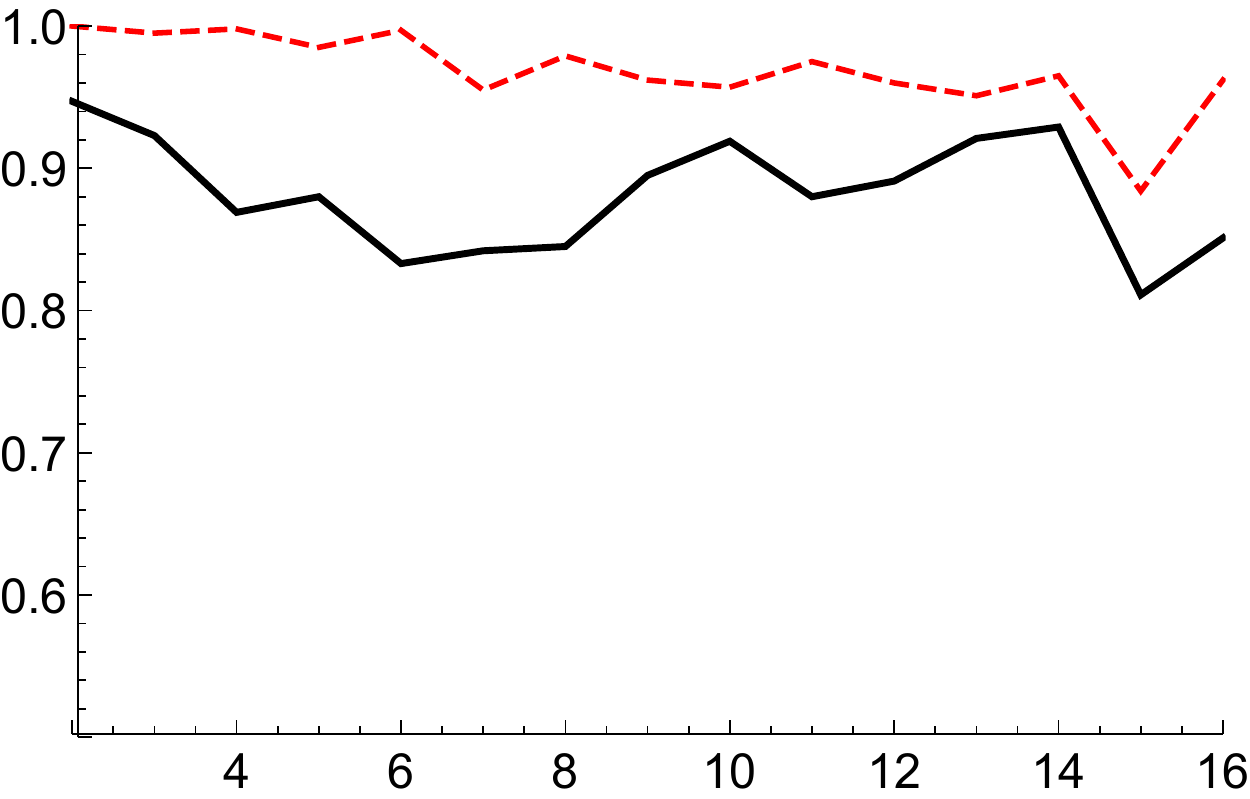}
	\caption{Success rate of the proper classification as the function of the
	number of elements in the learning set for the separable (black solid line)
	and non-separable (red dashed line) representation of data.}
	\label{fig:succ-rates-test1}
\end{figure}

The results of the described experiment are presented in
Fig.~\ref{fig:succ-rates-test1}. As one can observe the utilization of the
non-separable data representation allows better classification of the test data.
This is expected as the utilization of the tensor product enables the operations
on the larger subspace for the purpose of encoding the quantization symbols. 

%%%%%%%%%%%%%%%%%%%%%%%%%%%%%%%%%%%%%%%%%%%%%%%%%%%%%%%%%%%%%%%%%%%%%%%%%%%%%%%%
\section{Concluding remarks and further work}\label{sec:final}
%%%%%%%%%%%%%%%%%%%%%%%%%%%%%%%%%%%%%%%%%%%%%%%%%%%%%%%%%%%%%%%%%%%%%%%%%%%%%%%%
In the presented work we have introduced the complete quantum-inspired scheme
enabling the processing of real-world data for the purpose of pattern
classification. The introduced representations of data are based on the
quantization procedure which reduces the dimensionality of the quantum state
space required to process the data. Thanks to this one can implement and execute
the introduced scheme on a standard desktop computer. We have also used the
introduced framework to conduct a simple experiment where it is possible to
observe the difference between the separable and non-separable representation of
data.

One should note that the presented considerations are based on the formalism of
quantum mechanics, but do not require a working quantum computer. The
utilization of a quantum machine would be beneficial from the security point of
view as in this case the information about the class representatives is unknown
to the party performing the classification.

The introduced framework is based on a quantization procedure and the
introduction of the linear combinations of quantization symbols. As such it can
be easily adopted for other problems in machine learning. In particular we plan
to apply it to the problem of clustering. 

From the quantum information point of view, the introduced framework can be
generalized by using a formalism of mixed states in the quantization procedure.
One of the advantages of such approach is the possibility of using other
distance measures, based on the generalization of distance measures used in
classical information theory. Moreover, the use of density matrices can be used
for incorporating the information about uncertainty in the measurement results.

\subsubsection*{Acknowledgments.} This work has been supported by RAS project
on: "Modeling the uncertainty: quantum theory and imaging processing", LR
7/8/2007. The author would like to thank P.~Gawron, P.~Glomb, G. Sergioli and
L. Didacci for helpful discussions.

%\bibliographystyle{splncs03}
%\bibliography{../qpatterns,../qdiscrimination,../machine_learning}

\end{document}